\pdfoutput=1

\documentclass[useAMS,usenatbib]{mn2e}

\usepackage{upgreek}
\usepackage{bm}
\usepackage{amsmath}
\usepackage{epsf}
\usepackage{color}
\usepackage{natbib}
\usepackage{graphicx}
\usepackage{hyperref,ifthen}

\def\be{\begin{equation}}
\def\ee{\end{equation}}
\def\ba{\begin{eqnarray}}
\def\ea{\end{eqnarray}}
\def\nn{\nonumber}
\def\eqdef{\equiv}

\def\eprinttmp@#1arXiv:#2 [#3]#4@{
\ifthenelse{\equal{#3}{x}}{\href{http://arxiv.org/abs/#1}{#1}}{\href{http://arxiv.org/abs/#2}{arXiv:#2} [#3]}
}

\newcommand{\eprint}[1]{\eprinttmp@#1arXiv: [x]@}

\newcommand{\llponemul}[1]{\Upxi_{#1}}

\def\p{\Phi}

\def\a{{\cal A}}

\def\ta{\hat{\Theta}}
\def\tf{\bar{\Theta}}

\def\tu{\Theta}
\def\ctta{C^{\ta \ta}}
\def\cttl{C^{\tilde{\Theta} \tilde{\Theta}}}
\def\cttu{C^{\Theta \Theta}}
\def\cttn{C^{nn}}

\newcommand{\ThreeJ}[6]{\left(
                           \begin{array}{ccc}
        \! #1\! & #2\!  & #3\!  \\
        \! #4\! & #5\!  & #6\!
                           \end{array}
                   \right)}

\title[Lensing reconstruction from PLANCK sky maps: inhomogeneous noise.]{Lensing reconstruction from PLANCK sky maps: inhomogeneous noise.}
\author[Duncan Hanson and Graca Rocha and Krzysztof G\'orski]{Duncan Hanson$^{1}$ and Graca Rocha $^{2,3}$ and Krzysztof G\'orski $^{2,4}$\\
$^{1}$ Institute of Astronomy and Kavli Institute for Cosmology Cambridge, University of Cambridge, Madingley Road, Cambridge CB3 OHA \\
$^{2}$ Jet Propulsion Laboratory, California Institute of Technology, 4800 Oak Grove Drive, Pasadena CA 91109, U.~S.~A. \\
$^{3}$ Department of Physics, California Institute of Technology, Pasadena, 91125, U.~S.~A. \\
$^{4}$ Warsaw University Observatory, Aleje Ujazdowskie 4, 00478 Warszawa, Poland}
\begin{document}

\date{26 Aug 2009} % revtex4.
\pagerange{\pageref{firstpage}--\pageref{lastpage}} \pubyear{2009}  
\maketitle %mnras
\label{firstpage} %mnras

\begin{abstract}
\baselineskip 11pt
We discuss the effects of inhomogeneous 
sky-coverage on CMB lens reconstruction, 
focusing on application to the recently launched 
Planck satellite. We discuss the ``mean-field''
which is induced by noise inhomogeneities, as
well as three approaches to lens reconstruction in
this context: an optimal maximum-likelihood
approach which is computationally expensive to evaluate,
and two suboptimal approaches which are less intensive.
The first of these is only sub-optimal at the five per-cent level
for Planck, and the second prevents biasing due to uncertainties in the noise model.
\end{abstract}


% \maketitle revtex4

\section{Introduction}
The current generation of Cosmic Microwave Background (CMB) data has proven
to be remarkably well approximated as a statistically isotropic Gaussian random field 
\citep{Komatsu:2008hk}. 
Upcoming experiments are expected to push decisively past this approximation, 
to reveal a CMB which has been subtly distorted by gravitational lensing due 
to the large-scale-structure (LSS) which intercedes between ourselves and the 
surface of last scattering \citep{Lewis:2006fu}.
Mathematically, the effect of a fixed LSS realization is to make the CMB statistically 
anisotropic, introducing off-diagonal elements into its covariance. 
This enables one to construct estimators for the lensing potential
\citep{Hirata:2002jy,Okamoto:2003zw}. The power spectrum of the measured lensing potential
may then be used to obtain improved parameter constraints, particularly for
parameters which affect the late-time evolution of the Universe. The recently launched 
Planck satellite, for example, is expected to measure the CMB lensing signal internally with
cosmologically useful precision, enabling it to constrain the 
sum of neutrino masses to $0.1$eV \citep{Lesgourgues:2005yv}. Lensing is also important
as a potential contaminant for non-Gaussianity studies. The cross-correlation
between the lensing potential and ISW/Rees-Sciama induced temperature fluctuations
results in a bispectrum which has large overlap with the ``local'' type of non-Gaussianity.
Lensing therefore results in a bias to primordial non-Gaussianity estimation, which will
be significant for Planck \citep{Serra:2008wc}. Correction for this bias will be aided by
an accurate lensing reconstruction.

CMB lensing reconstruction works on the assumption that the underlying
CMB is statistically isotropic, and that any statistical anisotropy is
due to gravitational lensing. As such, it is potentially contaminated by
any systematic which introduces anisotropy onto the observed sky: 
beam asymmetries, astrophysical foregrounds, and inhomogeneous sky-coverage
are all expected to complicate the lens reconstruction.
In the absence of computationally expensive deconvolution mapmaking \citep{Armitage:2004pk}, 
beam asymmetries represent an unavoidable source of systematic error, which will need
to be quantified for any ultimate lensing analysis with Planck. 
Foregrounds, on
the other hand, may be cleaned to a high degree of accuracy by exploiting Planck's
wide frequency coverage. The magnitude of residual foregrounds at the small 
scales of interest to lensing reconstruction will not be adequately understood
until Planck has started to collect data, however. 
In this work, we will focus on
the effects of inhomogeneous sky-coverage. The scan strategy of Planck will
result in 
%strongly declination dependent sky coverage, 
noise levels which depend strongly on ecliptic latitude,
and so the effects of noise 
inhomogeneities are a large concern.
We will study both the optimal treatment of noise inhomogeneities,
as well as two suboptimal approaches: one which is computationally simpler than the optimal
reconstruction, and one which is insensitive to the instrumental noise model.

\subsection{Lens Reconstruction}
\label{sec:opt_ml}
We begin by reviewing the methodology of lens reconstruction. Consider a data model given by
\be
\ta(\Omega) = \tu(\Omega + \nabla \p(\Omega)) + n(\Omega)
\ee
where $\ta$ is the observed CMB, $\Omega$ picks out a location on the unit sphere, $\p$ is the lensing potential (for more details see \citealt{Lewis:2006fu}), $\tu$ is the primary, unlensed CMB with covariance $\cttu$, and $n(\Omega)$ is the instrumental noise realization with covariance matrix $\cttn$. 
Throughout this work, we will use a fixed flat $\Lambda$CDM cosmology for $\cttu$,
with standard parameters $\{ \Omega_b, \Omega_c, h, n_s, \tau, A_s \} \nolinebreak=\nolinebreak \{ 0.05, 0.23, 0.7, 0.96, 0.08, 2.4\times 10^{-9} \}$,
which is consistent with the WMAP5 best-fit power spectrum \citep{Nolta:2008ih}.
The maximum-likelihood estimator for the CMB lensing
potential in the limit of small $\p$ is given by 
\be
\hat{\p}_{LM} = \sum_{l'm'} \a_{LM, l'm'} [\tilde{\p}_{l'm'} - \langle \tilde{\p}_{l'm'} \rangle].
\ee
where $\a$ is a normalization matrix, the average is taken over realizations of the CMB and noise, and the un-normalized estimator $\tilde{\p}$ is given in harmonic space by
\be
\tilde{\p}_{LM}
= \frac{1}{2} \sum_{lm, l'm'} (-1)^{M} \ThreeJ{l}{l'}{L}{m}{m'}{-M} f_{l L l'} \tf_{lm} \tf_{l'm'}.
\ee
Here $\tf = (\ctta)^{-1} \ta = (\cttu + \cttn)^{-1} \ta$ is the inverse-variance filtered sky-map.
The indices $L$ and $M$ give the mode of the lensing potential which is being reconstructed.
The geometric term $f_{l L l'}$ is given (with the notation  $\llponemul{l} \eqdef l^2 + l$) by
\ba
f_{l L l'} &=&
\sqrt{\frac{(2L+1)(2l+1)(2l'+1)}{16\pi}} \ThreeJ{l}{l'}{L}{0}{0}{0} \nonumber \\
&& \left[ \cttu_{l}(\llponemul{L} + \llponemul{l} - \llponemul{l'}) + \cttu_{l'}(\llponemul{L} - \llponemul{l} + \llponemul{l'}) \right].
\ea
The estimator normalization matrix $\a$ is also equal to the estimator covariance $N^{\p\p}$. 
For more details, see \citealt{Hirata:2002jy, Smith:2007rg}.

This likelihood motivated estimator is closely related to the minimum-variance
quadratic estimator of Okamoto and Hu \citep{Okamoto:2003zw}, which may be derived 
under the assumption of uniform sky-coverage (homogeneous noise and no masking), in which
case $\cttn$ is diagonal. 
The likelihood approach motivates
two modifications which improve the performance of the estimator for non-uniform sky-coverage:
\begin{itemize}
\item{} 
Mean field subtraction: The effect of non-uniform sky coverage
is to introduce off-diagonal elements into the harmonic space 
noise covariance matrix which are interpreted by the estimator as lensing effects 
and give the estimator a non-zero expectation even in the 
absence of lensing.
\item{} 
Anisotropic inverse-variance filtering: This is an intuitive 
generalization from the 
%symmetric
rotationally invariant
 filters of the Okamoto and Hu estimator, although
difficult to derive in that context.
\end{itemize}
Similar ingredients are seen e.g. in bispectrum estimation \citep{Creminelli:2005hu}. 
Application of $(\ctta)^{-1}$ is generally a challenging problem at Planck resolution \citep{Smith:2007rg},
however with full-sky coverage we find that it is sufficiently well conditioned that it may be applied using 
conjugate-gradient descent with a diagonal preconditioner in less than one hour on a 2GHz processor to $\ell_{\rm max}=2500$, with an average 
fractional error of ${\cal O}(10^{-6})$ for each mode of the inverse-variance filtered field. 

As a baseline, we will also consider approximating the inverse variance filter as 
%symmetric, 
rotationally invariant,
taking only its diagonal elements, averaged over the azimuthal index $m$. 
In this case, we take $\ctta = \cttu + {\rm diag}( C^{nn} )$, where the diagonal operation is given by
\be
{\rm diag}( C )_{lm,\ l'm'}  = \delta_{ll'} \delta_{mm'} \frac{1}{2l+1}\sum_{m''} C_{l m'', l m''}
%= \frac{4\pi}{B_{\ell}^{2} n_{\rm pix}^2} \sum_{i} N_{i},
\label{eq:cl_nn}
\ee
We will refer to this as the ``uniform'' estimator,
and symbolically denote it in lowercase as $\hat{\phi}$.
The corresponding normalization and covariance matrices will accordingly be denoted as $A$ and $N^{\phi\phi}$ respectively. 
Expressions which involve $\phi$ and $A$ will implicitly be taken to use the symmetrized inverse variance filter as well.
The normalization to lens fluctuations is 
diagonal, independent of $M$, and may
%diagonal and may then 
be calculated analytically \citep{Okamoto:2003zw}:
\be
A_L = (2L+1)\left[ \sum_{l, l'} \frac{ f^2_{l L l'} }{2 \ctta_{l} \ctta_{l'} } \right]^{-1}.
\label{eq:al_unif}
\ee
With this normalization and accurate mean-field subtraction, the uniform estimator produces an unbiased reconstruction of the CMB,
however the normalization and the estimator variance are no longer explicitly equal (although we will see that in practice they are still very close).

\subsection{Noise Model}
\label{sec:noise_model}
To illustrate our discussion of inhomogeneous noise effects on lens reconstruction 
we will work with a semi-realistic model for Planck using simulated data from the 
detectors at $143$GHz, with an isotropic Gaussian beam of $\sigma_{\rm FWHM} = 7'$,
at HEALPix $N_{\text side}=2048$. 

We will assume that the map noise is Gaussian and effectively
 uncorrelated between pixels (prior to beam deconvolution).
In reality, non-white noise below the instrumental $1/f$ knee frequency leads
to inter-pixel noise correlations in the Planck data. To accurately study these
effects in the context of lens reconstruction requires many simulations generated
by performing the mapmaking procedure on realistic time-ordered data,
a computationally expensive task.
We leave this study to a future work, concentrating for now on pixel-uncorrelated noise.
In this case, the noise is completely characterized by a variance map. 
We obtain this map from the output of the \textit{Springtide} 
destriper mapmaker applied to a single realistic Planck simulation 
\citep{Ashdown:2007ta, Ashdown:2007tb}. 
The  noise levels which result have a white power spectrum which is twenty per-cent greater than that 
for the ``Bluebook'' value of $43\mu K \cdot {\rm arcmin}$ \citep{BLUEBOOK}, 
resulting in a cosmic variance limit of approximately $\ell=1500$. 
This enhancement of the white-noise noise level is due to the inhomogeneity
of the sky coverage.

The noise variance map is displayed in the upper panel of Fig.~\ref{fig:diag_filt_mf}.
The features in the noise variance are due primarily to the Planck scan strategy.
From the $L_{2}$ point of the Earth-Sun system, Planck spins at approximately
one rotation per minute about the anti-Sun direction. 
The angle between the optical axis and the spin axis is $85^{\circ}$,
and so the detectors trace rings which are nearly great circles on the sky. 
This results in relatively low noise levels near the ecliptic poles, where
there are many observations as well as a large degree of cross-linking. 
To gain sky coverage at the poles themselves, the Planck model which we have used
employs the so-called ``cycloidal'' scan strategy, in which the 
spin axis inscribes a circular path around the anti-Sun direction with a period 
of six months, keeping the angle between the spin axis and the anti-Sun direction 
at $7.5^{\circ}$. This results in the cusp features near the ecliptic poles in 
the upper panel of Fig.~\ref{fig:diag_filt_mf}. The thin circle of low-noise levels 
which connects the ecliptic poles is because the simulation which was used as
input for the mapmaking procedure included just over one year of data, and so
this section of the sky has on average fifty per-cent more hits than other regions. 

\section{Results}
\subsection{Mean field}

\begin{figure}
\begin{center}
\includegraphics[width=8.5truecm]{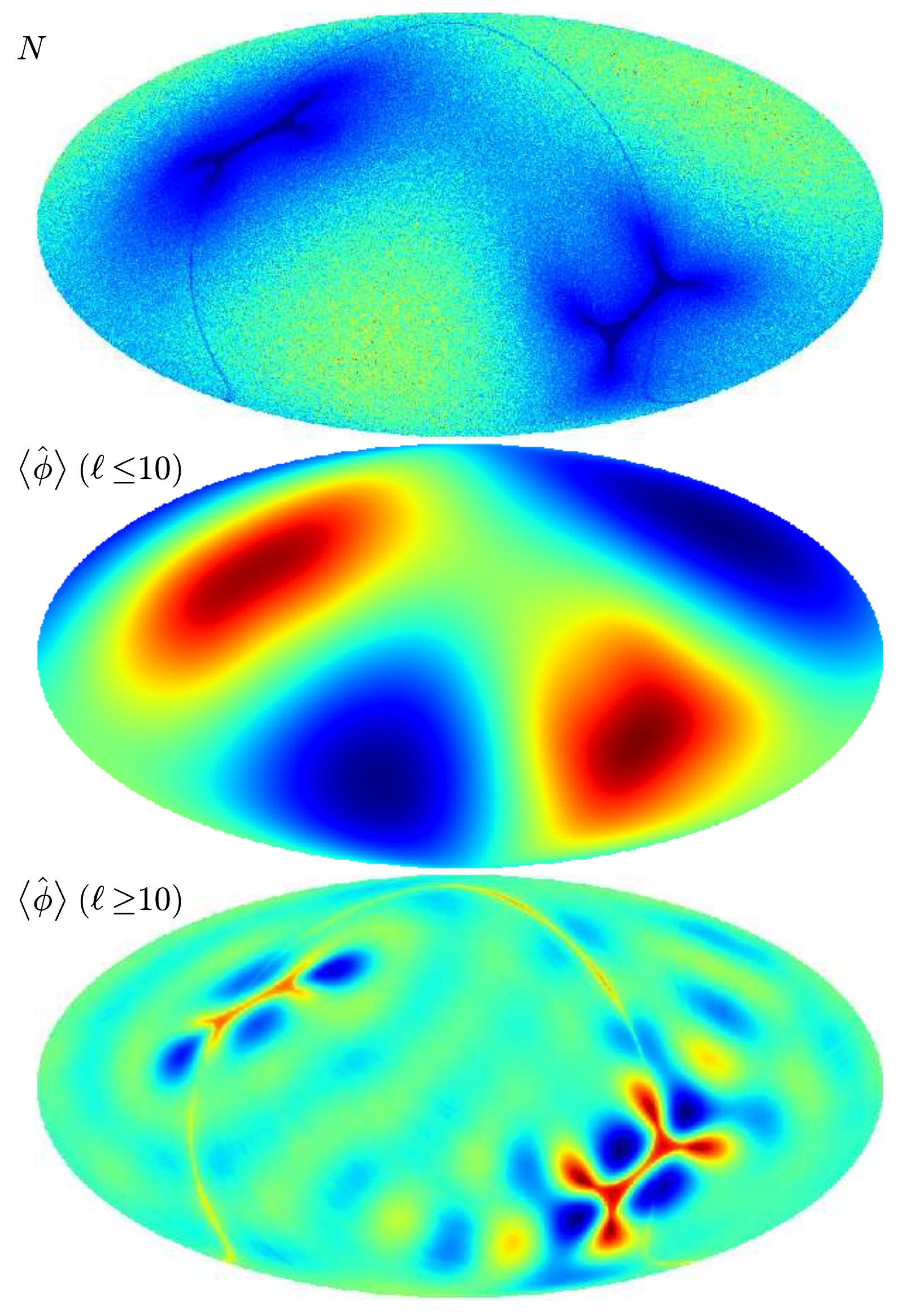}
\end{center}
\caption{Noise variance map (top) and lens reconstruction mean field (lower two panels), for our Planck noise model, in galactic coordinates. The thin stripe of low noise levels through the ecliptic plane is because we have used a cosmology in which data formats dominate over aesthetics at $z=0$.}
\label{fig:diag_filt_mf}
\end{figure}

\begin{figure}
\begin{center}
\includegraphics[width=8.5truecm]{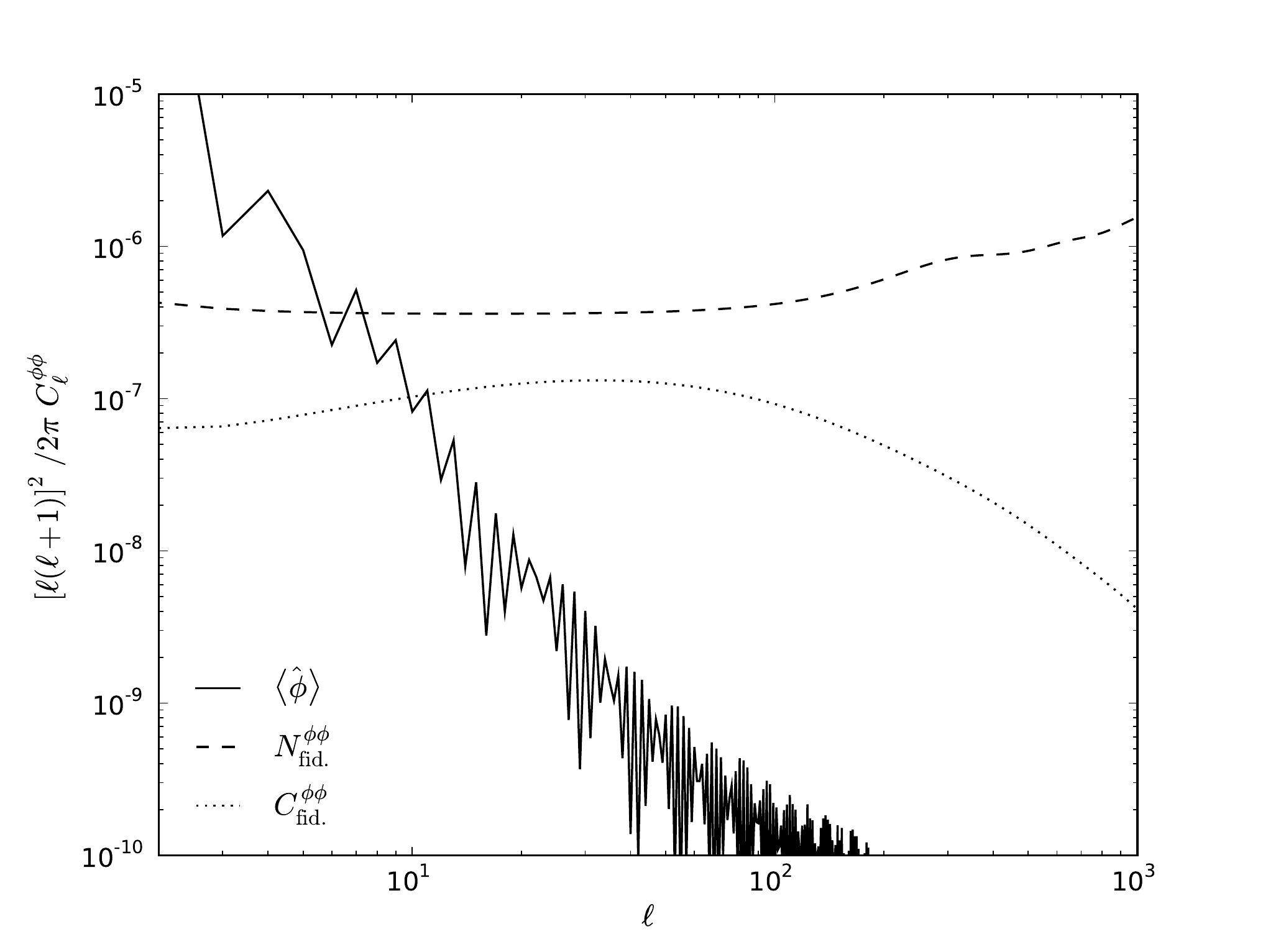}
\end{center}
\caption{The power spectrum of the mean field due to inhomogeneous noise for the uniform estimator (solid), the fiducial estimator variance $N^{\phi\phi}_{\rm fid.}$ (dashed), and the expected cosmological power spectrum of the lensing potential (dotted). }
\label{fig:mean_field_ps}
\end{figure}

We begin by considering the mean field term.
For the uniform estimator, the corresponding mean field may be calculated analytically:
\be
\langle \tilde{\phi} \rangle_{LM} = 
\sum_{lm, l'm'} (-1)^{M} \ThreeJ{l}{l'}{L}{m}{m'}{-M} f_{l L l'} \frac{ \cttn_{lm,\ l'm'} }{2 \ctta_{l} \ctta_{l'} }.
\label{eq:pmf_gen}
\ee
This expression may be reduced using the Gaunt integral and 
the orthogonality properties of the Wigner-$3j$ symbols to give a 
simplified expression for the mean field:
\ba
\langle \tilde{\phi} \rangle_{LM} &=&
N_{LM} \Bigg[ \sum_{l l'} \ThreeJ{l}{l'}{L}{0}{0}{0}^2 \frac{(2l+1) (2l'+1)}{n_{\rm pix} B_{l} B_{l'}} \nonumber \\
&& \quad \quad \frac{ \cttu_{l} [L(L+1) + l(l+1) - l'(l'+1)]}{2 \ctta_{l} \ctta_{l'}} \Bigg],
\ea
where $N_{LM}$ 
is the harmonic transform of the noise-variance map, $n_{\rm pix}$ is the number of map pixels and $B_l$ is the instrumental beam transfer function. We can see that for the uniform estimator, the mean field is simply the noise 
variance map convolved with 
%a symmetric 
a rotationally invariant
response filter.
In Fig.~\ref{fig:diag_filt_mf} we plot maps of the mean field for our Planck noise model. 
In Fig.~\ref{fig:mean_field_ps} we plot the power spectrum of the mean field, as well the expected 
lensing power spectrum and estimator variance. 
At low multipoles the magnitude of the mean field is considerably 
larger than the estimator variance for uniform noise. 

For the anisotropic estimator, the mean field term is 
even larger. In addition to the noise-only mean field, the anisotropic filtering
generates a mean field from the CMB anisotropies themselves.
%The response filter is very red, which leads to large low-multipole power in the mean field.
We will now consider the variance of the lensing reconstruction after subtraction of this large mean.

\subsection{Estimator variance}
\label{sec:estimator_variance}

For homogeneous white noise 
with a power spectrum given by Eq.~(\ref{eq:cl_nn}), the uniform estimator variance is equal to its normalization,
given in Eq.~(\ref{eq:al_unif}). We will therefore refer to this as the fiducial estimator noise, $N^{\phi\phi}_{\rm fid.} = A$. 

\begin{figure}
\begin{center}
\includegraphics[width=8.5truecm]{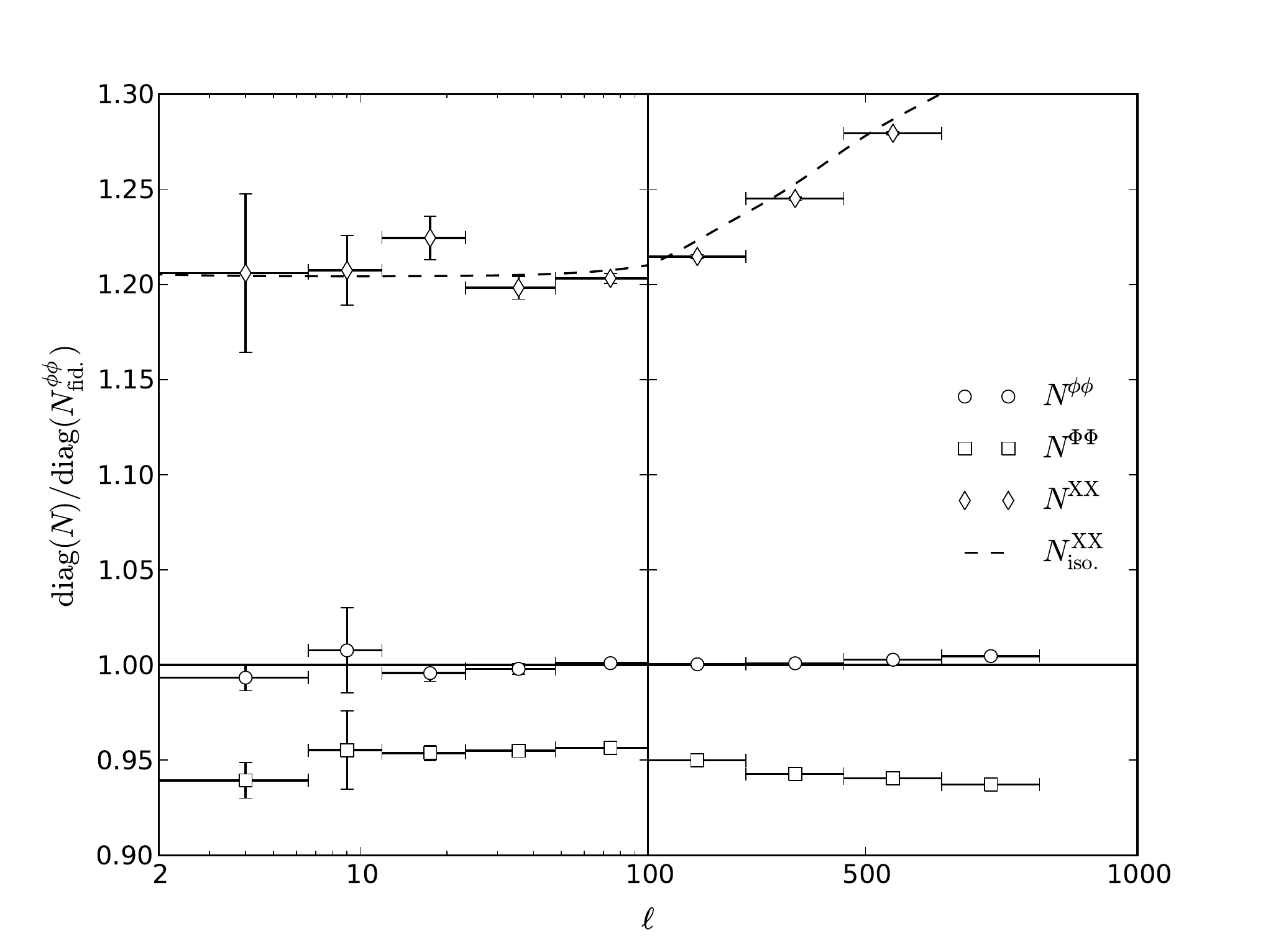}
\end{center}
\caption{Variance of lensing reconstruction with three different estimators, compared to the fiducial noise level: the ``uniform'' estimator (circles), the 
lower-limit of Eq.~\eqref{eqn:lower_limit} for the 
optimal ``anisotropic'' inverse-variance filter 
estimator (squares) and 
%the 
a pessimistic result for
cross-maps estimator (diamonds). Each point is the average for 100 Monte-Carlo simulations.}
\label{fig:variance_results}
\end{figure}

In Fig.~\ref{fig:variance_results} we compare the variance of the mean-subtracted uniform estimator
with this fiducial value. 
They agree well within our Monte-Carlo error bars for $\ell<300$.
The explanation for this agreement comes from considering the anisotropic, Gaussian noise distribution 
instead as an isotropic, non-Gaussian field, 
e.g. by randomizing the orientation of the scan strategy. 
For the uniform estimator, the isotropic component of the variance (equal to $N^{\phi\phi}_{\rm fid.}$)
is unaffected after mean-field subtraction, however the anisotropy of the noise distribution before averaging over orientation
 results in non-Gaussian connected terms. The ``primary'' configuration of this trispectrum
is removed by subtraction of the noise mean field, and at low-$\ell$ the remaining
configurations are suppressed by the estimator, as noted in \citealt{Okamoto:2002ik} and \citealt{Hanson:2027}.

For the anisotropic estimator, it is generally most efficient to compute the estimator covariance by Monte-Carlo, using
\be
{\rm Cov}[ \tilde{\p} - \langle \tilde{\p} \rangle ] = ( N^{\p\p} )^{-1},
\ee
keeping in mind that in our notation, the $\tilde{\p}$ estimates are un-normalized.
For a small number of parameters, estimation of this covariance matrix and subsequent inversion is typically stable. 
In the bispectrum context, for example, one usually seeks to determine the projection of a prescribed 
bispectrum shape in the data, a single parameter. 
In the case of lens reconstruction, however, the 
Fisher matrix will contain thousands of useful modes and is effectively impossible to obtain with sufficient accuracy from Monte-Carlo. 
In practice, this may not be an issue as many uses of the reconstructed lensing potential require 
$( N^{\p\p} )^{-1} \hat{\p}$ rather than $\hat{\p}$ itself (e.g. \citealt{Smith:2007rg}). In this work, we simply place a lower limit 
on the variance, using the result that
\be
{\rm diag}(N^{\p\p}) \ge \left[ {\rm diag}((N^{\p\p})^{-1}) \right]^{-1},
\label{eqn:lower_limit}
\ee 
which holds for any covariance matrix. 
This estimate of $N^{\p\p}$ is also compared to the fiducial variance in Fig.~\ref{fig:variance_results}. 
In practice the correlations between the reconstructed modes, which have been completely neglected here,
 reduce the amount of information in the estimator. The optimal anisotropic filtering results in a lensing estimator 
with approximately five per-cent less variance than the uniform estimator, a potentially useful improvement, 
although it must be kept in mind that this is an optimistic result.

At low multipoles we have seen that the magnitude of the mean field is considerably 
larger than the estimator variance. If the noise is well-understood, however, 
then the increase in the variance of the mean-subtracted estimates is negligible,
for both the optimal anisotropic and uniform filtering approaches. 
As a possible systematic check, in the next section we consider a lensing estimator which is 
insensitive to noise inhomogeneities.

\subsection{Cross-maps}
In this section we consider the behaviour of the lensing estimator on sets of
 maps with independent noise realizations. From the likelihood approach, given 
a set of CMB maps with independent noise the ``optimal'' treatment is to 
condense them to an inverse variance weighted average, which is then analysed 
as a single map (see e.g. \citealt{Hamimeche:2008ai}). To avoid biasing due to uncertainties 
in the noise model, it can be useful instead to work strictly with pairs of maps, 
such that auto-correlations of the noise are never produced. 
In the context of 
lens reconstruction, this allows us to avoid the mean field due to noise anisotropies. 
Given two noisy maps $a_{l m},\ b_{l m}$, we consider the cross-estimator
\be
\hat{\phi}^{a \times b}_{LM} =
\frac{A_L^{a \times b}}{2} \sum_{lm, l'm'} (-1)^{M}  \ThreeJ{l}{l'}{L}{m}{m'}{-M} f_{l L l'} \frac{{a}_{lm}}{D^{a}_l} \frac{b_{l'm'}}{D^{b}_{l'}},
\label{eq:phat_diag_x}
\ee
where the $D_{l}$ are isotropic filters and the normalization to lens fluctuations is  
%a straightforward generalization of the uniform estimator normalization given in Eq.~(\ref{eq:al_unif}):
given by
\be
A_L^{a \times b} = (2L+1)\left[ \sum_{l, l'} \frac{ f^2_{l L l'} }{2 D^{a}_{l} D^{b}_{l'} } \right]^{-1}.
\label{eqn:alx}
\ee
This type of estimator has been used by \cite{Hirata:2008cb} for the purpose of cross-correlating
$\hat{\phi}$ with large-scale structure from galaxy surveys. 
Here we are more interested in the lensing potential power spectrum,
which can be estimated as
\be
\hat{C}^{\phi \phi}_{L} = \frac{1}{2L+1} \sum_{M}
\frac{1}{|S|}
\sum_{(a,b)(c,d) \in S} (\hat{\phi}_{LM}^{a \times b}) (\hat{\phi}_{LM}^{c \times d})^{*},
\ee
where $S$ is a collection of map quadruplets and $|S|$ is its size. 
For simplicity we will assume that all of the maps have the same 
noise properties (although different noise realizations) and so the 
$D_{l}$ filters should all be equal.
In this case the reconstruction noise bias to $\hat{C}^{\phi \phi}$  
for homogeneous noise would be given by
\ba
N^{XX}_{\rm iso.} 
&=& \left(\frac{A_L^{X}}{2|S|} \right)^2 \Bigg[ \sum_{l l'} \frac{f_{l L l'}}{D_{l}^2 D_{l'}^2 } \nn \\
& & \quad \quad \quad \quad \sum_{(a,b)(c,d) \in S} (C^{ac}_{l} C^{bd}_{l'} + C^{ad}_{l} C^{bc}_{l'}) \Bigg],
\ea
where $A_L^X$ is Eq.~\eqref{eqn:alx} evaluated for the common $D_{l}$ filter. 
Noise anisotropies will manifest themselves in additional contributions to $N^{XX}$,
however we have already seen in Section~\ref{sec:estimator_variance} that these terms are small at low-$\ell$.
The choice of $S$ determines how sensitive this estimator is to
instrumental noise and inhomogeneities thereof. 

If we have $n$ independent maps and $S$ is taken to 
contain the ${}_4 {\rm C}_{n}$ quadruplets for which $(a\!\ne\!b\!\ne\!c\!\ne\!d)$, 
then it can be shown that $N^{XX}_{\rm iso.}$ is minimized for
$D_{l} = \cttl_{l}$ and is equal to the reconstruction noise level for 
a cosmic-variance limited experiment. This is directly analogous to the cross-correlation
approach which is often used for traditional power spectrum estimation, as it removes
any instrumental noise bias from the $C_{l}^{\phi \phi}$ estimates. 
The low reconstruction variance is somewhat misleading however, as the variance of the
power estimates contains contributions from the instrumental
noise which must be accounted for in a parameter analysis.
This approach discards a fraction $n! / [(n-4)! n^4]$ of
the possible map combinations, and so its effective sensitivity
must be less than the optimal approach of performing reconstruction on 
the minimum-variance sum of the maps, particularly for small $n$.

If we take $S$ to be the set of all quadruplets with $(a\!\ne\!b)(c\!\ne\!d)$ then
we retain more of the possible map combinations.
$\hat{C}^{\phi \phi}_L$ is no longer completely free of noise dependence, however 
the noise only enters $N_{\rm iso.}^{XX}$ through the map power spectra, which
are experimental ``observables'' and do not rely on any modelling of the noise.
In Fig.~\ref{fig:variance_results} we plot the variance of the cross-estimator measured
from simulations, relative to the minimum-variance result for the pessimistic case of only two maps.
We generate realizations of $a_{lm}$ and $b_{lm}$ with twice the noise variance levels of the previous simulations, 
such that the noise level of the minimum-variance average is unchanged. 
We inverse-variance filter them with $D_l = \cttl_l + 2\cttn$, which minimizes the term with
the largest contribution to $N_{\rm iso.}^{XX}$. 
If the $D$ filter were permitted to couple $l, l'$ then an estimator
with smaller variance could be derived, however it does not have
a known fast position space form to make its calculation feasible at Planck
resolution, and so we do not consider it here. In any case, our purpose in this 
section is to demonstrate a consistency test rather than a minimum-variance
reconstruction of $\phi$.
The sub-optimality of this approach is evident, with the variance of these estimates 
being approximately twenty per-cent larger than the fiducial value
(although this discrepancy will be less for a larger number of maps). 
The need to perform mean field subtraction has been obviated, however. 
The effect of the inhomogeneous noise is negligible at the level of the estimator variance in this approach. Only the noise and CMB power spectrum are required. Noise inhomogeneities do make contributions to the higher moments of the reconstruction statistics, however, as can be seen from the increased size of the Monte-Carlo error bars.

\section{Conclusions}

We have studied the effects of inhomogeneous 
instrumental noise on CMB lens reconstruction.
The main effect is to introduce a mean field even
in the absence of lensing, with large power on scales
$\ell < 100$.

We have studied the optimal estimator in the case
of such inhomogeneities, and found that for Planck it
performs approximately five per-cent better than a
suboptimal approach which is computationally less 
expensive and easier to study analytically. 
With accurate mean field subtraction, the suboptimal ``uniform'' estimator 
itself performs as well as would be expected for homogeneous noise
with the same power spectrum.

Both the optimal and uniform reconstructions require an 
accurate modeling of the noise inhomogeneities to be effective. 
This requirement may be bypassed using a lens reconstruction based
on pairs of maps with uncorrelated noise, which we suggest will
provide a useful consistency test.

\section{Acknowledgments}
Some of the results in this paper have been derived using the HEALPix \citep{Gorski:2004by} package. DH is grateful for the support of a Gates scholarship, and to Anthony Challinor for useful discussion.
We gratefully acknowledge support  by the National Aeronautics and Space Administration (NASA) Science Mission Directorate via the US Planck Project. 
The research described in this paper was partially
carried out at the Jet Propulsion Laboratory, California Institute of
Technology, under a contract with NASA. 

\bibliographystyle{mn2e_eprint}
\bibliography{lensing_with_noise}

\begin{thebibliography}{}

\bibitem[\protect\citeauthoryear{Armitage \& Wandelt}{Armitage \&
  Wandelt}{2004}]{Armitage:2004pk}
Armitage C.,  Wandelt B.~D.,  2004, Phys. Rev., D70, 123007,
  \eprint{astro-ph/0410092}

\bibitem[\protect\citeauthoryear{Ashdown et~al.,}{Ashdown
  et~al.}{007a}]{Ashdown:2007ta}
Ashdown M. A.~J.,  et~al., 2007a, A\&A, 467, 761, \eprint{astro-ph/0606348}

\bibitem[\protect\citeauthoryear{Ashdown et~al.,}{Ashdown
  et~al.}{007b}]{Ashdown:2007tb}
Ashdown M. A.~J.,  et~al., 2007b, A\&A, 471, 361, \eprint{astro-ph/0702483}

\bibitem[\protect\citeauthoryear{Creminelli, Nicolis, Senatore, Tegmark \&
  Zaldarriaga}{Creminelli et~al.}{2006}]{Creminelli:2005hu}
Creminelli P.,  Nicolis A.,  Senatore L.,  Tegmark M.,    Zaldarriaga M.,
  2006, JCAP, 0605, 004, \eprint{astro-ph/0509029}

\bibitem[\protect\citeauthoryear{Efstathiou, Lawrence \& Tauber}{Efstathiou
  et~al.}{2006}]{BLUEBOOK}
Efstathiou G.,  Lawrence C.,    Tauber J.,  2006, \eprint{astro-ph/0604069}

\bibitem[\protect\citeauthoryear{Gorski et~al.,}{Gorski
  et~al.}{2005}]{Gorski:2004by}
Gorski K.~M.,  et~al., 2005, ApJ, 622, 759, \eprint{astro-ph/0409513}

\bibitem[\protect\citeauthoryear{Hamimeche \& Lewis}{Hamimeche \&
  Lewis}{2008}]{Hamimeche:2008ai}
Hamimeche S.,  Lewis A.,  2008, Phys. Rev., D77, 103013, \eprint{0801.0554}

\bibitem[\protect\citeauthoryear{Hanson, Challinor, Efstathiou \&
  Bielewicz}{Hanson et~al.}{2009}]{Hanson:2027}
Hanson D.,  Challinor A.,  Efstathiou G.,    Bielewicz P.,  2009, In prep.

\bibitem[\protect\citeauthoryear{Hirata, Ho, Padmanabhan, Seljak \&
  Bahcall}{Hirata et~al.}{2008}]{Hirata:2008cb}
Hirata C.~M.,  Ho S.,  Padmanabhan N.,  Seljak U.,    Bahcall N.~A.,  2008,
  Phys. Rev., D78, 043520, \eprint{0801.0644}

\bibitem[\protect\citeauthoryear{Hirata \& Seljak}{Hirata \&
  Seljak}{2003}]{Hirata:2002jy}
Hirata C.~M.,  Seljak U.,  2003, Phys. Rev., D67, 043001,
  \eprint{astro-ph/0209489}

\bibitem[\protect\citeauthoryear{Komatsu et~al.,}{Komatsu
  et~al.}{2009}]{Komatsu:2008hk}
Komatsu E.,  et~al., 2009, ApJS, 180, 330, \eprint{0803.0547}

\bibitem[\protect\citeauthoryear{Lesgourgues, Perotto, Pastor \&
  Piat}{Lesgourgues et~al.}{2006}]{Lesgourgues:2005yv}
Lesgourgues J.,  Perotto L.,  Pastor S.,    Piat M.,  2006, Phys. Rev., D73,
  045021, \eprint{astro-ph/0511735}

\bibitem[\protect\citeauthoryear{Lewis \& Challinor}{Lewis \&
  Challinor}{2006}]{Lewis:2006fu}
Lewis A.,  Challinor A.,  2006, Phys. Rept., 429, 1, \eprint{astro-ph/0601594}

\bibitem[\protect\citeauthoryear{Nolta et~al.,}{Nolta
  et~al.}{2009}]{Nolta:2008ih}
Nolta M.~R.,  et~al., 2009, ApJS, 180, 296, \eprint{0803.0593}

\bibitem[\protect\citeauthoryear{Okamoto \& Hu}{Okamoto \&
  Hu}{2002}]{Okamoto:2002ik}
Okamoto T.,  Hu W.,  2002, Phys. Rev., D66, 063008, \eprint{astro-ph/0206155}

\bibitem[\protect\citeauthoryear{Okamoto \& Hu}{Okamoto \&
  Hu}{2003}]{Okamoto:2003zw}
Okamoto T.,  Hu W.,  2003, Phys. Rev., D67, 083002, \eprint{astro-ph/0301031}

\bibitem[\protect\citeauthoryear{Serra \& Cooray}{Serra \&
  Cooray}{2008}]{Serra:2008wc}
Serra P.,  Cooray A.,  2008, Phys. Rev., D77, 107305, \eprint{0801.3276}

\bibitem[\protect\citeauthoryear{Smith, Zahn \& Dore}{Smith
  et~al.}{2007}]{Smith:2007rg}
Smith K.~M.,  Zahn O.,    Dore O.,  2007, Phys. Rev., D76, 043510,
  \eprint{0705.3980}

\end{thebibliography}

\label{lastpage}
\end{document}